# Helping Blind People in Their Meeting Locations to Find Each Other Using RFID Technology

Farshid Sahba
Institute for Informatics and Automation Problems
National Academy of Sciences of Armenia
Yerevan, Armenia
f_sahba@yahoo.com

Amin Sahba, Ramin Sahba
Department of Electrical and Computer Engineering
The University of Texas at San Antonio
San Antonio, TX, USA
{amin.sahba, ramin.sahba}@utsa.edu

*Abstract*— **This paper presents a new specific system based on RFID technology to help blind people find the other party in their meeting place. This system, uses a device called Smart Director or SD, equipped with Active Tag and RFID Reader. The blind person, the visitor, adjusts the SD with the identification number of the other party who has shared it previously for example on a telephone call, and takes the device to the meeting place. When the blind person arrives at the location, the reader of the device receives signals from the active tag of the other party's SD. It then identifies his/her position based on the intensity and direction of the received signals and tell it to the blind person. In this way, blind people can simply identify the position of the other party in crowded places and find each other.**

## I. INTRODUCTION

The blind people are part of the society, and they are busy with their jobs and doing things daily just like other people. If blind people, like other people, are to benefit from environmental and urban services, we should consider their specific daily needs and design and supply appropriate tools to meet their needs.

One of these daily needs is to handle personal or work appointments in the city. The blind person goes to a location in the city for the meeting after coordinating the appointment with the other party in a telephone conversation. For a blind person it is challenging to find the other party who may have never met before, in a public place like a street, square, park, restaurant, coffee shop, or lounge, since he/she cannot observe their surroundings in that place and find the other party in the crowd. This becomes even more serious if the other party (the visitee) is also blind, so no one can see the other. Calling the name of the other party is not a solution, because the sound is lost in the crowd, or it may be prohibited to create a loud voice at the desired location.

Radio frequency identification (RFID) is one of the technologies that have being used for localization. RFID uses radio signals to identify objects and helps to detect existence and location of them [1,2]. Although newer technologies such as global positioning system (GPS) and Wi-Fi positioning system (WPS) are used today for positioning, the former is not effective for indoor localization, and the latter is dependent on the existence of WiFi in the building. Recently, there have been many researches using RFID for localization in specific situations and improving its efficiency. However, none of them has addressed localization issue for blind people.

Authors in [3] combine RFID technology into a navigation system with GPS and a gyroscope to improve the accuracy of positioning specifically in tunnels or downtown areas where GPS signal is weak or lost. In [4], authors propose a real time indoor RFID positioning system which uses passive tags to lower the cost as an economical solution for identifying the location and movement of personnel and goods. Researchers in [5] present a new positioning algorithm which uses two mobile RFID readers, and passive or active randomly distributed tags with known locations. The location of the unknown tag then can be estimated by using the multilateration equation with the locations of the known tags and applying Kalman Filtering technique. The work proposed in [6] is an RFID positioning system that uses a Kalman filter and calculates the location of the object based on signal strength disseminated from nearby RFID tags and a database including tag numbers and their positions. Authors in [7] present two hybrid indoor positioning methods which both are based on wireless sensor network (WSN) and RFID. One approach uses an extended Kalman filter (EKF) technique whereas the other one exploits a particle filter (PF) algorithm. A new technique is proposed in [8] to accurately estimate the indoor location of pedestrians by combining inertial navigation system (INS) and active RFID technology. This approach uses received signal strengths (RSSs) from multiple active RFID tags with known locations to help a position estimation method based on foot-mounted inertial measuring units (IMUs). In [9], authors propose an RFID-based approach for indoor tracking and positioning of small-sized laboratory animals. Their method exploits a near-field (NF) RFID multiantenna system which works in the ultra high frequency (UHF) bandwidth, to identify and locate the passive NF RFID tags implanted in animals. They also develop and integrate an algorithm based on the measured received signal strength indication (RSSI) to enhance the accuracy of the tracking system. Authors in [10] present a portable RFID indoor positioning system which uses one RFID reader as the targeted device and two active RFID tags as the landmarks. Their proposed system utilizes Kalman-filter drift removal (DR) and Heron-bilateration location estimation (LE) techniques. Presented approaches in [11,12] suggest that providing each object with multiple tags significantly improves the efficiency of RFID systems when encountering radio noise, object occlusions, and other interfering factors. [13-15] use RFID technology for automation and control in some smart systems.

## II. PRESENTED MODEL

In order to overcome the daily challenges of visual impairment, the blind people use simple and useful tools such as a white cane to sophisticated and expensive tools such as sight prostheses. All these tools are functional, and a blind person usually chooses one or more tools depending on their life, environmental, physical, and financial conditions to overcome those challenges. Variety and innovation in providing these tools can make life easier for the blind people and increase their productivity. A well-designed model is required for an applied system [16-22], and it should explain all features of the system [23-26]. This paper presents design of a practical system along with a device whereby the blind people can easily find each other at public and perhaps crowded places for their appointments. Although the concept behind the system is not so complex, it can address complicated problems for its users [27-35].

The system is based on RFID technology, and the device comes with a battery, an active tag, an RFID reader, a receiver and transmitter antenna, a microcontroller, a keyboard, a speaker, and a vibrator.

A unique 9-digit number is stored within the active tag, and the tag is broadcasting this number continuously in the environment through the radio waves using transmitter antenna. This number is also engraved on the body of the device on Braille, so the owner of the device knows the unique number of it.

This device is called the Smart Director, which will be referred to briefly as the SD in this article. Each of the two parties (the visitor and the visitee) must have an SD device.

The microcontroller memory (ROM) has two applications, which is explained below.

After the meeting location is determined in a phone conversation and receiving a unique SD number from the other party, the first program is executed by the blind person before going to the meeting location, and the unique SD number of the other party is entered to the program to be stored in the memory. From now on, the SD becomes sensitive to this number and if the SD is placed in an environment where this number is in the radio waves of that location, then the second program, which is discussed next, runs. The pseudocode of this algorithm is shown in Figure 2.

```
void Save_ID()
{
    cin>> ID_Code;
    if (Is_Valid(ID_Code))
    {
            Store(ID_Code, DataBase);
            Beep;
    }
    else
    {
        Beep;
        Beep;
    }
}
```

Fig. 2. Pseudocode of the algorithm for storing the unique SD number of the other party in the memory.

The second program identifies the direction of the visitor's location based on the data received from the SD reader, and, with the command given to the vibrator, determines the vibration intensity. So, the visitor will be informed about the direction of the location of the visitee by the vibration. The pseudocode of this algorithm is shown in Figure 3.

```
void Person_Detector()
{
    while (1)
    {
        cin>> ID_Code;
        cin>> Wave_Power;
        if (ID_Code == Restore(DataBase))
            ON_Vibra (Wave_Power);
    }
}
```

Fig. 3. Pseudocode of the algorithm for identifying the direction of the visitor's location based on the data received from the SD reader

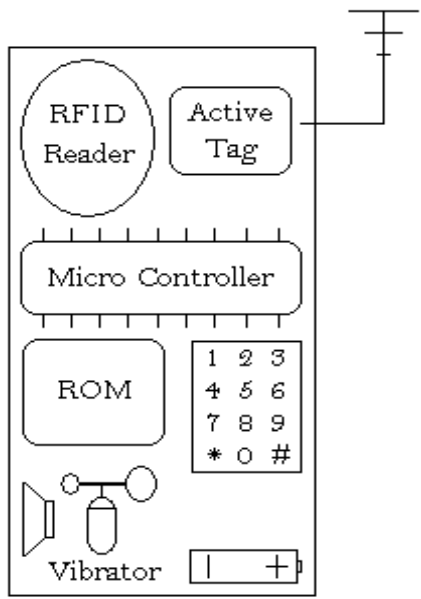

Fig. 1. Smart Director (SD)

As previously said, the blind person receives the unique number of the SD device of the other party when they are setting up the time and location of the appointment over the phone or using other communication devices, and then enters that number into their SD device after the call. At this point, he can turn off the device and turn it on when arriving at the location of the appointment. At that time, the scenario of finding the standing location of the other party should be run. This scenario is represented by the activity diagram in Figure 4.

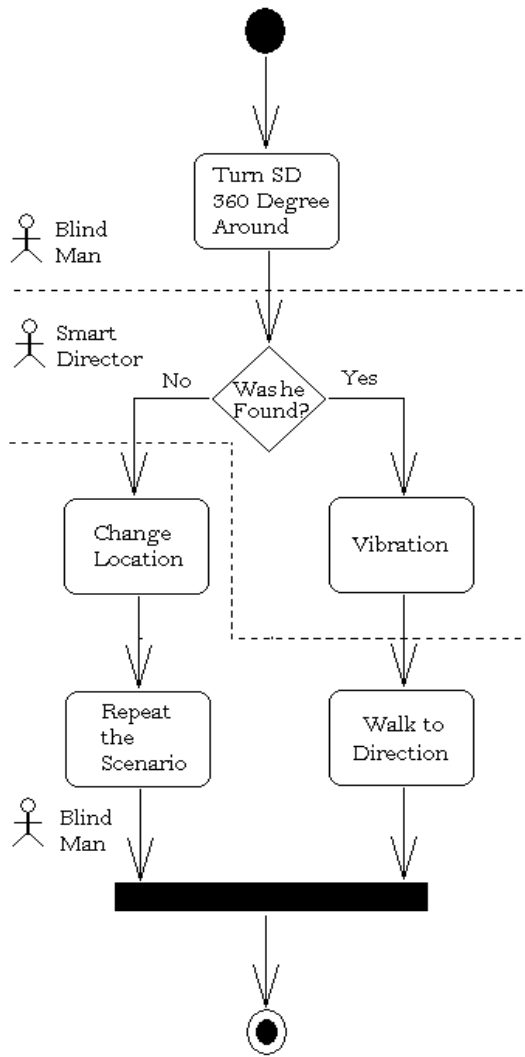

Fig. 4. The activity diagram for the scenario of finding the standing location of the other party at the location of the appointment.

By running the aforementioned scenario, or repeating it if needed, the blind person will find if the other party is at an average distance of 50 meters in the surroundings, and steps in the direction that vibrator of the SD warns to reach the exact position of the other party. It needs to be mentioned that the visitee shouldn't change his/her position during this process or it will take a longer time for the blind person to find the location of the other party.

### III. EXPERIMENT AND RESULTS

To test the functionality of the system, its circuit was implemented on a bread board in the lab of Rayan Shid Nama engineering company (RSN). A monitor was used instead of the vibrator in the circuit. Figure 5 below shows the implemented circuit.

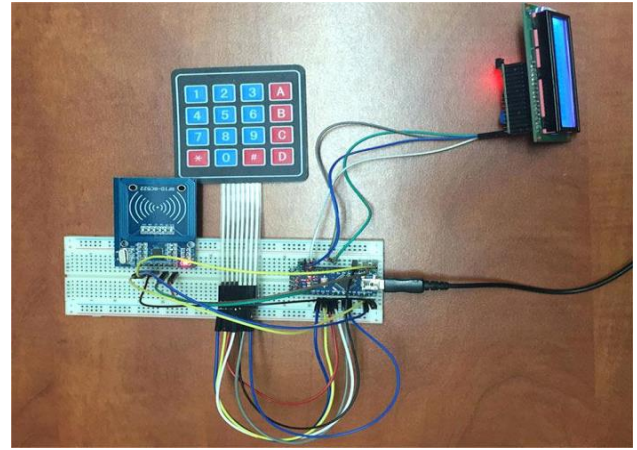

Fig. 5. The implemented circuit of the system

The active RFID tag was prepared in a wrist strap and placed on the hand of one of the staff members in the role of the visitee which is shown in Figure 6.

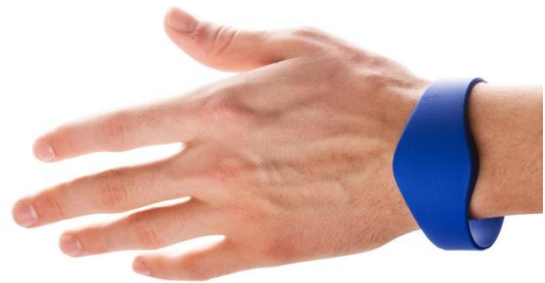

Fig. 6. RFID tag in a wrist strap

Afterwards, the person with the wristband stood as a visitee in a totally dark hall with an approximate area of 500 square meters. Another staff member played the role of a blind person and entered the hall as a visitor while carrying the circuit with himself. He did not know the exact location of the first person (visitee) in the hall.

After entering, he rotated the circuit around 360 degrees, in any direction. If the message received by the signal on the display, he moved in the same direction, and if not, he changed the place of standing randomly and repeated rotating the circuit 360 degrees around until he could reach to the location of the visitee. The experiment was repeated 16 times, and each time the visitee stood in a new location in the hall. At each experiment, the spend time for the visitor to reach to the location of the visitee was measured and noted. The average of this time was 86 seconds.

In another experiment in the same dark hall, the visitee stand again in a new location, and then a recorded sound of an environment in one of the squares of Tehran was played at 65 dB. Then the visitor entered the room with no equipment and tried to find the location of the visitee by calling her name and hearing her answer until reaching her location. This experiment also was repeated 16 times, and the average time for the visitor to find the location of the visitee was 340 seconds.

## IV. Conclusion and future work

By comparing the results of the two experiments, it can be seen that the use of the proposed system and the Smart Director reduces the time taken to find the location of the visitee by 74% for the appointments of the blind people in urban areas. It is expected that this proportion of time optimization in large urban public places will have a significant impact on reducing the time costs for the blind people and bring about relaxation for them in these situations.

For a definitive assessment, this system can be piloted with the participation of several blind volunteers and the planning of appointments between them in urban public places.

The strengths of this system are the simplicity of training and deployment for the blind people, the cheapness of equipment and software, and its high efficiency. However, the weaknesses of this system are the need to provide its equipment and software for each blind person, the weight and load gain of blind person's mobile devices when going to an appointment, and also the need to rotate the SD around 360 degrees when looking for the visitee.

In order to overcome the weaknesses of the system, a smartphone and its RF waves, Bluetooth or Wi-Fi can be used instead of the SD, because the smartphone has all the hardware components of the SD inside its own. The software can also be developed as a mobile app and installed on the smartphone. In this way, providing the equipment and software is simplified and the weight and volume of blind person's mobile devices are reduced. If smart glasses are used instead of SD, then the visitor should only turn their head around when looking for the visitee, which is much more common and easier than turning the SD around by hand. Another solution to eliminate the need to rotate the SD 360 degrees is installing four RF receiver antenna in four sides of the SD.

As a recommendation to develop this system, it is possible to equip the proposed system in this article with a GPS receiver, so that the blind person will be guided by the GPS system from the beginning of the route to the meeting location, and then find the visitee with the help of the RF of the SD device in the meeting location.

Another suggestion of extending this system is to use it for appointments in very crowded, dark, or less known locations. The receiver antenna and the software of the SD can also be upgraded to increase the accuracy of its functionality, so that, in addition to the direction, the distance to the visitee is estimated and notified to the visitor in real time.